\documentclass[letterpaper,onecolumn]{IEEEtran}

\usepackage[utf8]{inputenc} 
\usepackage[T1]{fontenc}    
\usepackage{hyperref}       
\usepackage{url}            
\usepackage{booktabs}       
\usepackage{amsfonts}       
\usepackage{nicefrac}       
\usepackage{microtype}      
\usepackage{lipsum}
\usepackage{graphicx}

\title{Understanding Important Features of Deep Learning Models for Transmission Electron Microscopy Image Segmentation}

\author{\IEEEauthorblockN{James P. Horwath\IEEEauthorrefmark{1}, Dmitri N. Zakharov\IEEEauthorrefmark{2}, Remi Megret\IEEEauthorrefmark{3}, Eric A. Stach\IEEEauthorrefmark{1}}
\\
\IEEEauthorblockA{\IEEEauthorrefmark{1}Department of Materials Science and Engineering, University of Pennsylvania, Philadelphia, PA}
\\
\IEEEauthorblockA{\IEEEauthorrefmark{2}Center for Functional Nanomaterials, Brookhaven National Laboratory, Upton, NY}
\\
\IEEEauthorblockA{\IEEEauthorrefmark{3}Department of Computer Science, University of Puerto Rico, Rio Piedras, San Juan, PR}}

\begin{document}

\maketitle

\begin{abstract}
Cutting edge deep learning techniques allow for image segmentation with great speed and accuracy. However, application to problems in materials science is often difficult since these complex models may have difficultly learning physical parameters.  In situ electron microscopy provides a clear platform for utilizing automated image analysis. In this work we consider the case of studying coarsening dynamics in supported nanoparticles, which is important for understanding e.g. the degradation of industrial catalysts.  By systematically studying dataset preparation, neural network architecture, and accuracy evaluation, we describe important considerations in applying deep learning to physical applications, where generalizable and convincing models are required.
\end{abstract}


\section{Introduction}
In situ  and operando experimental techniques, where dynamic process can be observed with high temporal and spatial resolution, have allowed scientists to observe chemical reactions, interfacial phenomena, and mass transport processes to give not only a better understanding of the physics of materials phenomena, but also a view into how materials react under the conditions in which they are designed to perform\cite{Zheng2015,Tao_2017}.  As the use of in situ techniques continues to expand, and technology to enable these experiments continues to develop, we are faced with the fact that more data can be produced than can be feasibly analyzed by traditional methods\cite{Taheri2016,Hill2016}.  This is particularly true for in situ electron microscopy experiments, where high resolution images are captured at very high frame rates.  In practice, hundreds of images can be captured per second. However many experimental analyses consider less than one frame per second, or even one frame for every several minutes\cite{Simonsen}.  Methods for fast and efficient processing of high-resolution imaging data will allow for not only full utilization of existing and developing technologies, but also for producing results with more statistical insight based on the sheer volume of data being analyzed.

Simultaneously, the field of computer vision provides well understood tools for image processing, edge detection, and blob localization which are helpful for moving from raw image data to quantifiable material properties.  These techniques are easy to apply in many common computer programming languages and libraries. However more recent research highlights the processing speed and accuracy of results obtained through the use of machine learning\cite{Badea2016,Chen2014}.   Previously, a combination of traditional image processing and advanced statistical analysis has be shown to successfully segment medical images\cite{Dheeba2011,Bengio2015}.  Deep learning - generally using multi-layer neural network models - expands on other machine learning techniques by using complex connections between learned parameters, and the addition of non-linear activation functions, to achieve the ability to approximate nearly any type of function\cite{Goodfellow-et-al-2016}.  With regards to image segmentation and classification, the use of Convolutional Neural Networks (CNNs), in which high-dimensional learned kernels are applied across grouped image pixels, is widespread.  CNNs provide the benefit that their learned features are translationally invariant, meaning that image features can be recognized regardless of their position in the image.  This makes such models useful for processing images with multiple similar features, and robust against variation in position or imaging conditions\cite{Moen2019}.  Additionally, the feature richness of high-dimensional convolutional filters and the large number of connections between hidden layers in a neural network allows for the learning of features which are too complex to represent manually, and which make intuitive interpretation difficult.  Much of the literature studying CNNs focuses on high-accuracy segmentation/classification of large, complex, multi-class image datasets or upon improving data quality through super-resolution inference, rather than quantitative analysis of high-resolution images\cite{Yang2018}.  While additional memory requirements alone make processing of high-resolution images difficult, the scale of features and possible level of precision also changes as a function of image resolution.  Most importantly, for the simple case of particle edge detection, the boundary between classes in a high-resolution image may spread across several pixels, making segmentation difficult even by hand.  To our knowledge, no systematic study on the use of full high-resolution images (image size larger than 512 x 512 pixels) when training a CNN for semantic segmentation exists: generally, these images are broken into smaller parts or rescaled, requiring additional preprocessing steps and creating the potential to introduce artifacts.

Though, it seems, the tools for rapid segmentation of high-resolution imaging data exist, several points of concern regarding the use of deep learning must be acknowledged.  First, though the ease of implementation using common programming tools enables extension of methods to new applications by non-experts, the complexity and still-developing fundamental understanding of deep learning can lead to misinterpretation of results and poor reproducibility\cite{Zhang2017a,Wang2019}.  Moreover, models can be prone to overfitting - memorizing the data rather than learning important features - which can go unnoticed without careful error analysis\cite{Dietterich1995,Srivastava2014}.  Overfitting occurs when a model has enough parameters that an unrealistically complex function can be fit to match every point in a data set.  Thus, a model which accurately labels data by overfitting will likely fail when shown new data since its complex function does not describe the true variation in the data. Therefore, and overfitted model isn’t useful for future work.  Finally, the high dimensionality of data at intermediate layers of a neural network combined with the compound connections between hidden layers makes physical representation of learned features impossible without including more assumptions into the analysis.  These challenges – specifically representation and visualization of CNN models – are areas of active research\cite{Selvaraju2017,Umehara2019}.

Building on previous work on image segmentation, automated analysis, and merging deep learning within the field of materials science, we focus on the semantic segmentation of Environmental Transmission Electron Microscopy (ETEM) images of supported gold nanoparticles\cite{Madsen2018,Schneider2017,Ziatdinov2017,Zakharov2018}.  This allows efficient measurement and tracking of particles frame-to-frame, and thus will aid in the understanding of fundamental processes governing coarsening and ripening in industrial catalysts at the atomic scale\cite{Hansen2013,Ostwald1900}.  Here, we study a variety of CNN architectures to define the most important features for the practical application of deep learning to our task.  We discuss the impact of dataset preparation and collection, how image resolution affects segmentation accuracy, and the role of regularization and preprocessing in controlling model variance.  Further, we investigate how image features are learned, so that model architectures can be better designed depending on the task at hand. By using a simpler approach to semantic segmentation, in contrast to poorly understood and highly complex techniques, we intend to show that common tools can be utilized to construct models which are both accurate and physically meaningful.

\section{Results and Discussion}
\subsection{Convolutional Nerual Network Design and Performance on 512x512 Resolution Images}
Scientists in several fields, most notably medical imaging, have worked on semantic segmentation tasks similar to the one considered here\cite{Shen2017,Wilson2016}.  Previous results have shown that an encoder-decoder-type architecture is well suited for semantic segmentation tasks where both the spatial position and classification accuracy are vital on a pixel-by-pixel level.  With this approach, successively deeper convolutional/max-pooling layer pairs (added to decrease spatial resolution while simultaneously increasing feature richness) are combined with up-sampling convolutional layers that aim to re-scale the image back to a higher resolution while decreasing the feature dimension of the image source\cite{Ziatdinov2017,Badrinarayanan2017,Karianakis2015}.  Our experiments show that a network architecture of this type successfully segments many images at 512x512 resolution, however the centers of large particles are consistently missed, resulting in detected particles with a doughnut-like shape (as seen in Figure S2).  We attribute this problem to an inability of the CNN to learn similar features with different size-scales; we suspect that, if our model detects edges, the lack of variation in the interior of the particle appears similar to that in the background.  While translational invariance is often discussed as a key property of convolutional neural networks, CNNs are not scale invariant. This means that a model which successfully recognizes a feature with a small size may not necessarily recognize the same feature at a larger size; this has to do with the size of the receptive field - here, kernel size - of the convolutional layer relative to the size of the image. Moving towards CNN scale-invariance, scientists have used CNN architectures with multiple columns, each learning features of a different scale\cite{Kanazawa2014}.  We tested a simpler method for improving upon this short-coming, by increasing the size of the convolutional kernel to improve the model’s ability to capture multi-scale features.  This approach was successful, however increasing the kernel size further increases the number of learnable parameters and memory required, thereby increasing the complexity of the model and the training time required for the model to converge.

\begin{figure}[h]
    \centering
    \includegraphics{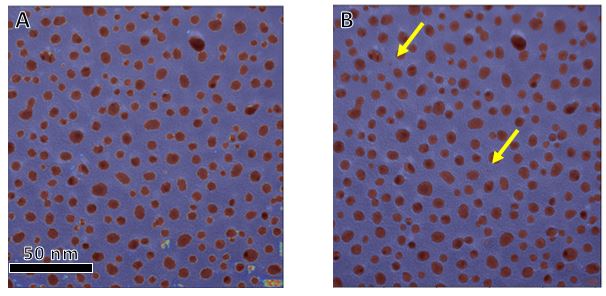}
    \caption{Segmentation results for UNet-type architecture on 512x512-resolution images. a.) shows raw output from the model overlaid on the raw image; notice the sharp activation cutoff at particle edges. b.) Threshold applied to image to show final segmentation result. Yellow arrows indicate that small particles were successfully recognized in contrast to earlier trials.}
    \label{UNet_init_fig}
\end{figure}

Additionally, we tested a CNN architecture inspired by the UNet\cite{Ronneberger2015}.  This model, rather than increasing kernel size with the goal of expanding the receptive field, uses skip-connections to tie activations in the encoding stage to feature maps in the decoding stage in order to improve feature localization.  Skip connections work by concatenating encoded and decoded images of the same resolution followed by a single convolutional layer and activation to relate unique aspects of both images (see visual representation in Supplemental Figure S3).  This improves upon the similar encoder-decoder architecture, where images are scaled down and then back up to learn local features rather than pixel-level features, by maintaining local environments from the original image to map features in the output.  Results using the UNet-type architecture on our image set show that the model is able to consistently recognize both large and small particles, and that it is robust against varied imaging conditions and datasets (Figure \ref{UNet_init_fig} and S2, show results on images from experiments not represented in the training set). 

Each of our CNN outputs has a final softmax activation for binary classification on a pixel-by-pixel basis.  The mathematical action of a softmax function, comparing the activation of each class to the total activation for a given pixel, allows softmax output to be qualitatively related to the probability or confidence of the given pixel label.  As a probability density map, labelled images were transformed to a true binary classification by thresholding the particle activations at 0.7 (nominally, 70\% prediction accuracy).  

\subsection{Dataset Properties}
The necessity of a large amount of accurately labelled data is a major deterrent towards the use of machine learning techniques in many fields. This is only exacerbated when using deep learning, where the large number of parameters, and abstraction introduced by multiple layers of non-linear activation can easily cause overfitting on small datasets. While several data augmentation and sampling methods exist to expand the utility of a small number of manually labeled data points, semantic segmentation presents a particularly difficult task since it requires labels for individual pixels in an image rather than a single classification for an entire image.  In our case, a high imaging frame rate provides a large volume of potential training data, however images captured successively are not independent and look very similar. Leveraging the large number of particles present in a single image, independent training data can nevertheless be obtained by random crops of the appropriate size from images and by shifting images through an Affine Transformation.  Further augmentation would involve including rotation and scaling variation.

We find that pixels corresponding to particles make up only about 15\% of the training data.  Class imbalance may make learning image features difficult when using convolutional neural networks, however we see no obvious detrimental effects in our results.  The general feature sparsity of our raw images, in addition to the clear separation between nanoparticles, and the apparent similarity in nanoparticle shape and size in our images likely mean that fewer complex features must be learned to produce successful segmentation.  In other words, though particle-pixels are under-represented in the training data, they clearly stand out when seen in the local context of background-pixels.     

Our tests show that experimental factors such as blur, due to difficulty imaging thick samples or slight sample drift, or uneven illumination, due to aberrations in a condensed electron beam or poor voltage center alignment, have little effect on the output of deep learning models as long as these variations are accounted for in the training set. Along these lines, a CNN trained on noisy data is not able to identify particles in images with less noise, even though to the human eye particles are easier to find in cleaner images.  We tested this by passing averaged images (40 consecutive frames), which would have increased signal-to-noise ratio, through a CNN trained on images from the same experiment which were not averaged (Figure S4).  Particle detection in averaged images was hindered by the presence of imaging and experimental artifacts which were more apparent when the amount of random variation in the image was reduced.  This observation does not show overfitting, but rather illustrates the fact that averaged images essentially come from a different distribution than the noisy training data, since the same model was able to successfully segment images from a completely different experiment which showed a similar amount of noise to that in the training set. This additionally emphasizes that thinking of CNNs learning features of images as humans due can be misleading.

\subsection{Higher Resolution Image Segmentation}
While segmentation of 512x512-pixel images is possible and seemingly accurate, many advanced electron microscopes are equipped with cameras which allow even higher pixel resolutions.  The accuracy of manual particle measurements from images with different resolutions likely changes very little (assuming accuracy is mainly dependent on the care taken by person making measurements), however changes in resolution, particularly around particle edges, can greatly influence automated labeling performance which generally relies on edge contrast.  For an image with a fixed side length, increasing pixel resolution decreases the relative size of each pixel.  Decreasing the pixel size increases the possible measurement precision, and therefore, high-resolution images are needed to provide both accurate, and consistent particle measurements.  Along these lines, the error introduced by mislabeling a single pixel decreases as pixel density (image resolution) increases.

\begin{figure}[h]
    \centering
    \includegraphics{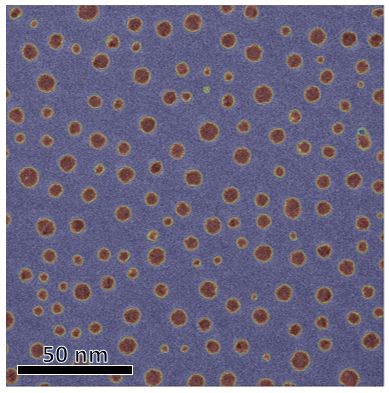}
    \caption{Using the same UNet architecture but increasing image resolution makes it more difficult for the model to localize edge features}
    \label{blurry_edges_fig}
\end{figure}

\begin{figure}[h]
    \centering
    \includegraphics{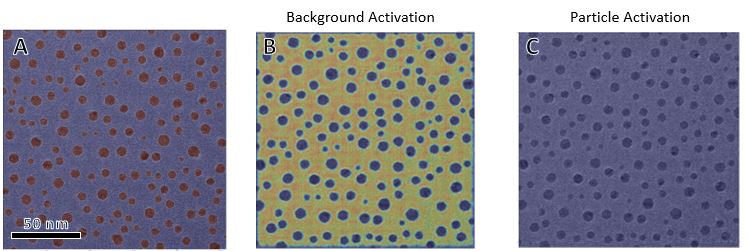}
    \caption{a.) shows the CNN output for a given image. b.) and c.) show the raw activation values for layers detecting background and particles, respectively.  The softmax function combines these activation maps to produce a.).}
    \label{no_activation_fig}
\end{figure}

A base, 3-step U-Net architecture, as described in the methods section, was trained on a dataset of 1024x1024 ETEM images, yet performance was significantly worse based qualitatively on edge sharpness and magnitude of softmax activation as seen in Figure \ref{blurry_edges_fig}.  While the previous results gave proper segmentation with a softmax threshold of 0.7 (a relatively high confidence), 1024x1024 images used a threshold of merely 0.4 to produce results similar to low-resolution models.  Moreover, we noticed that training the same model on the same data more than once would produce different results: while in some cases training produced image segmentation with wide edge variation, other training instances gave segmentation results with nearly perfectly identified particles, with little to no variation at particle edges.  These results likely signal overfitting of the dataset, with the model ‘memorizing’ the noise rather than actual features, as raw activation maps (Figure \ref{no_activation_fig}) show that in fact no features of particles are learned by the model and instead only noise patterns in the background areas are recognized.  This model, therefore, would likely produce very accurate particle measurement on the training dataset, yet would not generalize to data from other experiments or with particles of different sizes.  This is further highlighted by the instability of the model with respect to the length of training time.  Our experiments show that a model which may appear to be converged after some number of training epochs may revert back to poor performance before settling again.  Indeed, a model like the one in Figure \ref{no_activation_fig} would show a sudden drop followed by no further change in learning rate as a function of training time.  Though decreasing the learning rate could help to avoid this problem, this would only make it more difficult for our model to learn image features.

Rather than solely increasing the width and depth of the model to improve the performance (we used a 4-step UNet-type architecture for 1024x1024 images, see Supplemental Figure S3), we developed several hypotheses to describe which factors may cause the poor performance of the previously successful network with increased image resolution.  First, the developers of UNet explain that they included multiple convolutional layers at each level of down-/up-sampling with the assumption that it is successive convolutions across spatial and feature dimensions that help to combine previous layers and localized activations with spatial correlations from skip-connections\cite{Ronneberger2015,He2016}.  With this inspiration we add a second convolutional layer at each step to improve feature localization (see Figure S3).  One noticeable change in the raw images between 512- and 1024-resolution images is the sharpness of the contrast at particle boundaries. For lower resolutions, where pixels represent larger distances, the boundary between a particle and background can be captured in a single pixel, causing a sudden change in image contrast which would be easier to capture than a slow change over a large number of pixels.  Further increases in image resolution would make edge detection even more difficult.  Therefore, we investigated the effect of adding Gaussian blur to the raw images to reduce the relative impact of high frequency variation in the background and increase the impact of contrast changes at particle edges.  Further, we suspected that the inability of the model to learn edge characteristics may be due to a vanishing gradient which can arise when negative convolution output is pushed to a value of zero by the ReLU activation function. This is known as the dying ReLU problem and may be solved by replacing all ReLU activation instances with Leaky ReLU\cite{Goodfellow-et-al-2016,Maas2013}.  Finally, wide variation in raw activation values for particles within the same image led us to believe that our CNN had large variance.  We implemented batch normalization as a means to regularize the variation in activation at each convolutional layer in the network architecture\cite{Ioffe2015}.  The effect of learning rate on model performance was also investigated in order to determine how to best sample the loss landscape, but in this regard, we found that a learning rate of 0.0001 is practical and effective for all deep models on our dataset.

\begin{figure}[h]
    \centering
    \includegraphics{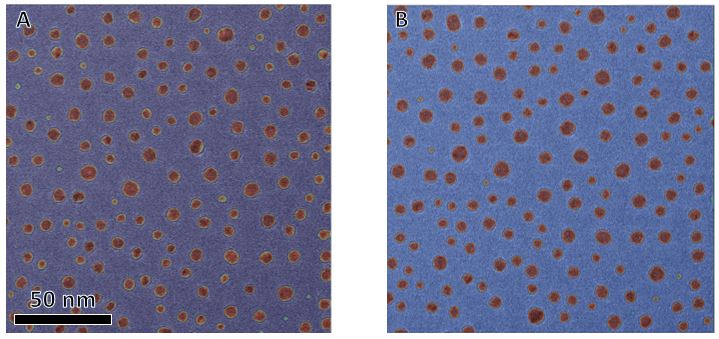}
    \caption{Best results, shown in a.), came from applying batch normalization at each convolutional layer, with no blue and a learning rate of 0.0001. b.) shows further imporvement by increasing the convolutional kernel size from 3x3 to 7x7.}
    \label{best_results_fig}
\end{figure}

Fifteen different CNN configurations based on the original UNet architecture were tested to study the influence of Gaussian blur with different $\sigma$ values, adding a second convolutional layer at each up-sampling step, batch normalization, and learning rate (either 0.001, or 0.0001).  Results from all fifteen models are shown in Supplemental Figure S5.  Model performance here is qualitatively based on whether particles of varying sizes were detected, sensitivity to noise and illumination variation in the raw image, and the sharpness of the activation cutoff at particle edges.  Based on these criteria, best performance is seen in models with batch normalization only (Figure \ref{best_results_fig}a), and batch normalization combined with extra convolutional layers.  In both cases models trained on both blurred ($\sigma = 2$) and unblurred images perform similarly on samples taken from the training set. From this, it appears that Batch Normalization is the most important factor for learning particle features from 1024x1024mages.  When training from scratch, i.e. without pretrained weights, it has been shown that the loss function is smoother and model convergence is better when using Batch Normalization, which may have a significant effect on higher resolution images due to the combinations of strong noise and lack of visually discriminative features at lower scales of analysis\cite{Santurkar2018}.  Our use of ReLU activation functions essentially produces output values in the range $[0,\infty)$, which presents a risk of divergence of the activation values, and can be mitigated by normalization in successive convolutional layers before the final softmax activation.  Leaky ReLU allows activations on the range $(-\infty,\infty)$, however the small activation for negative pixel values, combined with batch normalization, works to avoid increasing variance with the number of convolutional layers.  Aside from applying batch normalization, we find that the only way to achieve significant segmentation improvement on high resolution images is to increase the size of the convolutional kernel, here from 3x3 pixels to 7x7 (Figure \ref{best_results_fig}b).  However, this greatly increases the number of trainable parameters and training time for the model.

To briefly summarize the practical implications of our findings, continual Batch Normalization through successive convolutional layers has a significant positive effect on the performance.  For our dataset, increasing network depth does not appear to increase the performance of the CNN.  A slow learning rate produces the best results and most stable models, while preprocessing training images with Gaussian blur seems to increase the risk of overfitting.

\subsection{Evaluating Detection Accuracy}
With lower resolution images, softmax output, as a measure of detection confidence, could be used to threshold an image.  For the best of the fifteen models tested on the 1024x1024-pixel dataset, this method could only be used to segment images using a threshold value of 0.4.  Taken as a measure of confidence, these results are unfavorable.  However, considering this as a suitable threshold regardless of the interpretation gives consistent particle sizes through an image series which would provide measurements at least as usable as manually measured data.

\begin{figure}[h]
    \centering
    \includegraphics[angle=270,scale=0.75]{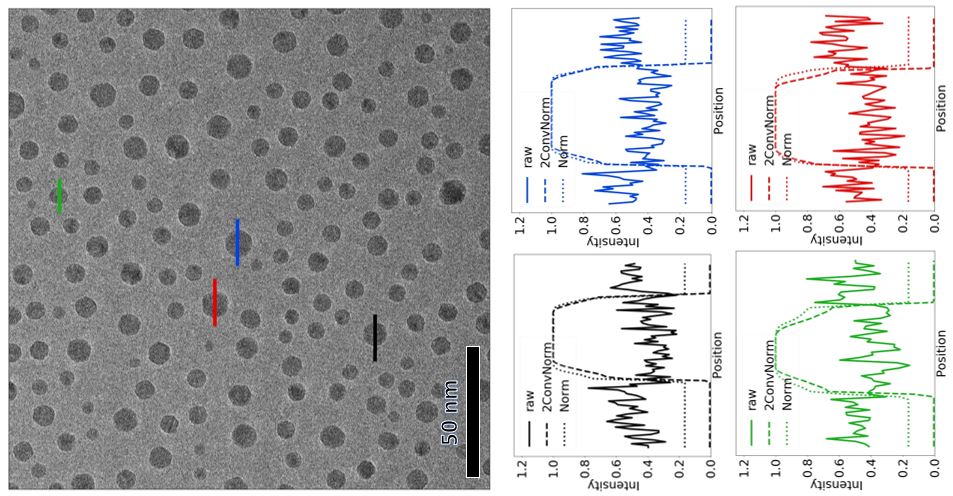}
    \caption{Intensity profiles for selected particles in a training image.  Line scans show intensity variation for each particle in the raw image (solid), network with batch normalization (dotted), and network with batch normalization and extra convolutional layers (dashed).}
    \label{line_profiles_fig}
\end{figure}

\begin{figure}[h]
    \centering
    \includegraphics[scale=0.75]{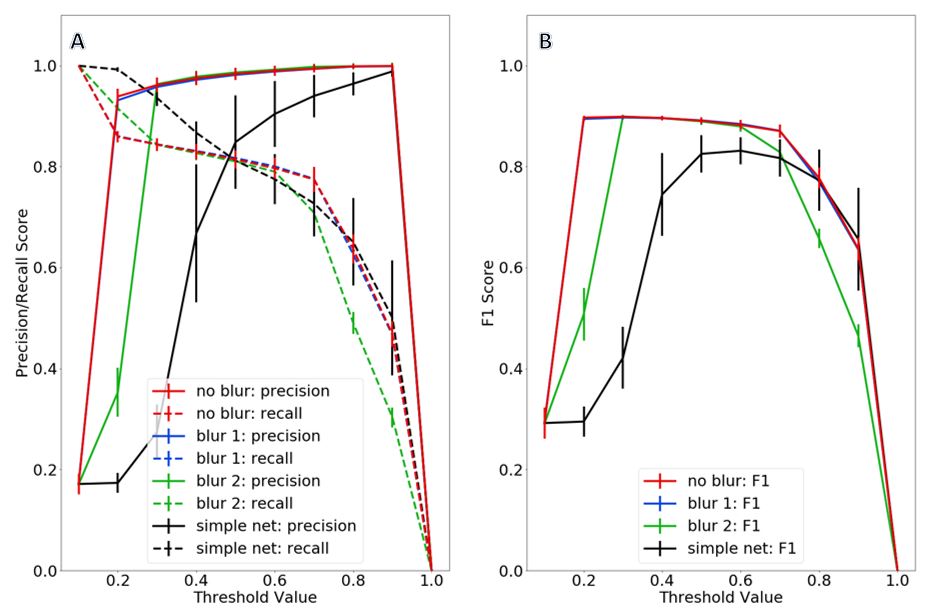}
    \caption{Accuracy metrics (a: precision and recall; b: F1 score) for UNet-type, and simple one-layer CNN architectures with added batch normalization presented as a function of the amount of blur applied to training data and chosen segmentation threshold.  Note that, in both a and b, the red no-blur and blue blur-1 curves almost completely overlap.}
    \label{accuracy_metrics_fig}
\end{figure}

Variation in the color scale at particle edges, as seen Figure \ref{blurry_edges_fig}, led us to believe that our particle measurement would greatly vary as a function of the chosen threshold.  Intensity line profiles, as shown in Figure \ref{line_profiles_fig} are helpful in illustrating this edge variation for two models compared to the intensity of the raw image.  These plots check how two different models perform in comparison with the edge contrast in the raw image.  Indeed, as the intensity approaches 1, both models show a slope in intensity.  Figure \ref{accuracy_metrics_fig} collects average precision, recall, and F1 accuracy scores for the batch-normalized CNN as a function of threshold value, and the amount of Gaussian blur applied compared to a set of 50 validation-set labels.  Here, high precision means that the model produces few false positives (pixels labeled as particle which actually correspond to background), while recall measures the proportion of particle pixels which were successfully identified by the model.  The F1 score is the harmonic mean of precision and recall.  Based on these results, we could expect that the normalized models with no applied blur and blur ($\sigma = 1$) are stable with respect to precision and recall at a particle activation threshold values below 0.7.  The model trained on blurred images with $\sigma = 2$, shows similar performance over a smaller range of stable thresholds.  For our case of binary classification of an unbalanced dataset, where recognizing particles pixels is more important than recognizing background, recall is likely the most important measure for determining a threshold for use in practice.  While we see convergence with maximum precision for the model without blur around a threshold of 0.7, we realize that our empirically selected value of 0.4 gives better recall with essentially the same precision as compared to thresholding at 0.7.  Our experience shows that the Otsu threshold, which separates the intensity histogram such that the intra-class variance is minimized, is a practical choice for segmenting our data\cite{Otsu1979}.  This makes sense, since, qualitatively, we see a large peak close to 0 activation representing the background with nearly all pixels with higher activation values corresponding to particles.  However, it can be shown mathematically that the calculated Otsu threshold may mislabel the class with a wider intensity distribution\cite{Xu2011}.  Therefore, thresholding datasets with a lower signal-to-noise ratio would likely be more difficult.  In these cases, it is imperative that a large dataset is used for training, as choosing low threshold values, even when they produce usable results, makes it difficult to recognize overfitting.

\subsection{Learning Features with a Simpler Model}
\begin{figure}[h]
    \centering
    \includegraphics{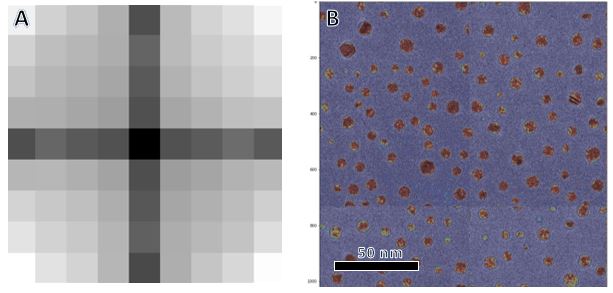}
    \caption{The kernel (a) learned by a single-layer CNN and the segmentation it produces (b, after softmax activation).}
    \label{learned_kernel_fig}
\end{figure}

An effective machine learning model requires a balance between the number of learnable parameters, the complexity of a model, and the amount of training data available in order to prevent over-fitting and ensure deep-learning efficiency\cite{Lu2017,Kabkab2016}.  In an efficient model, a vast majority of the weights are used, and vital to the output.  In practice though, deep networks generally have some amount of redundant or trivial weights\cite{Han2015}.  In addition to efficiency, several issues have come to light regarding the use of deep learning for physical tasks which require an interpretable and explainable model as this often leads to better reproducibility and results which generalize well\cite{Umehara2019,Kabkab2016}.  Rather than increasing model complexity to perfectly fit difficult or noisy data, it is better practice to refine assumptions and understanding of the raw data and develop a model to test these hypotheses.  With this philosophy in mind, we decided to test a significantly pared down CNN, with a single convolutional layer consisting of a single learnable filter followed by softmax activation on our training data which produced the segmentation shown in Figure \ref{learned_kernel_fig}b.  The benefit of such an architecture is that since the dimensionality of the kernel is the same as that of the image we can easily visualize the learned weights (Figure \ref{learned_kernel_fig}a).  This visualization can be interpreted in several ways.  First, we can conceive that the algorithm is learning vertical and horizontal lines (dark lines), potentially similar to a basic Gabor filter for edge detection – though it is missing the common oscillatory component - combined with some amount of radially-symmetric blur (light gray). Alternatively, we can envision that the horizontal/vertical lines could be an artifact of the electron camera or data augmentation method meaning that the learned filter represents an intensity spread similar to a Laplacian of Gaussian (LoG) filter which is used to detect blobs by highlighting image intensity contours.  As a simple test of our supposition, Supplemental Figure S5 shows that a sum of a horizonal Gabor filter, vertical Gabor filter, and Gaussian filter qualitatively produces a pattern similar to our learned kernel.  Previous work has similarly shown that edges and other spatially-evident image features are learned in the early convolutional layers of a CNN\cite{Yosinski2014}.   Repeating the same method with another kernel size, this time 7x7-pixels rather than the initial 9x9, produces a similar filter, showing that the results are not an artifact of the feature scale.  Such a single-layer model with logistic activation is analogous in practice to applying a linear support-vector machine (SVM) for linear regression\cite{Baudat2002}.

While this model is useful for illustrating the power of simpler machine learning methods, minimal changes are needed to extend this idea to a model that provides usable, practical segmentation.  Using one convolutional layer, now with 32 filters, followed by a second, 1x1 convolutional layer to combine the features into a segmented image, we test a shallow but wide CNN architecture.  Again, aside from the convolutional layer used to combine the extracted features, filters from this shallow network can be visualized to see what features are being learned from the data.  The precision, recall, and F1 scores of this simpler model (Figure \ref{accuracy_metrics_fig}, black line) are comparable to the performance of the deep networks described above.  These results illustrate that a model with significantly fewer parameters and quicker training time can still produce a usable segmentation.  Similarly, our results suggest that shallow, wide CNNs have enough expressive power to segment high resolution image data\cite{Lu2017}.  Though 32 filters (visualized in Figure S6) may be too many filters to directly compare for visually extracting useful information, it is possible to see a general trend: filters are learning faint curved edges.  This is not surprising, since in the simpler model with a single convolutional filter learns more general straight edges.  Moreover, taking the mean of all 32 filters (Supplemental Figure S7) shows a similar pattern as Figure \ref{learned_kernel_fig}a with slight rotation.  Further analysis of the set of 32 filters would require regularization of the entire set of weights to allow for more direct comparison.  For many semantic segmentation tasks useful in materials science, at least to first order, features are determined based on edge contrast.  Therefore, increasing model width, as opposed to depth would allow scientists to directly analyze the features of their data to evaluate if the trained model is effective, and whether it will generalize to other datasets.  Based on these results we expect that designing a shallower neural network which retains the local semantics learned in an encoder-decoder or UNet architecture would make a generalizable model for particle segmentation more realistic.  

\section{Conclusions}
We have systematically tested several design aspects of CNNs with the goal of evaluating deep learning as tool for segmentation of ETEM images.  With proper dataset preparation, the choice of a suitable learning rate, and continual regularization, standard CNN architectures can easily be adapted to our application.  While overfitting, class imbalance, and data availability are overarching challenges for the use of machine learning in materials science, we find that knowledge of data features and hypothesis-focused model design can still produce accurate and precise results.  Moreover, we demonstrate that meaningful features can be learned in a single convolutional layer, allowing us to move closer to a balance between state-of-the-art deep learning methods and physically meaningful results.  Indeed, we evaluate the accuracy of several deep and shallow CNN models and find evidence that, for a relatively simple segmentation task, important image features are learned in the initial convolutional layers.  We apply common accuracy measures to evaluate our models, but we note that other specially designed metrics may help to define exactly where mistakes are made, and thereby which features a model is unable to represent.

Comparing the filters learned by single-layer CNNs to common image processing filters suggests that simpler segmentation methods may be effective.    Moreover, particularly for cases like our high-resolution images where class boundaries are unclear and are continually obscured with increasing image resolution, we note that semantic understanding - where each pixel’s local surrounds are used to predict a segmentation value - is more important than extraction of information from complex, high-dimensional feature spaces.  For this reason, we believe that, whether shallow or deep, fully-convolutional architectures, which at each position consider a local receptive field, are superior to non-convolutional neural networks or methods which require manual feature extraction, such as common linear SVMs.

Additionally, we present a basic method for automated training set generation which may allow for easier application of supervised deep learning models to other datasets.  While our method works as a quick and efficient means of generating ground truth, there are several points of note for further image analysis studies.  First, further testing shows that using an active contour algorithm, compared to our use of filters and morphological reconstruction, produces more precise ground truth labels.  However, active-contour labeling requires an initial segmentation guess to provide accurate results, so this type of labeling would likely take a longer time.   Second, though our automated segmentation appears reliable on a dataset consisting of nanoparticles on a flat, homogenous substrate, stronger background variation or the presence of multiple object classes would make automated generation of training images more difficult.  For example, images of nanoparticles on a heterogenous support, in the presence of three-dimensional nanostructure, or with a multi-modal particle size distribution or varying nanoparticle chemistry would likely be much more difficult to segment.  Indeed, even automated segmentation using traditional computer vision approaches is difficult in these scenarios.

In all, while computer science research trends towards complicated, yet highly accurate deep learning models, we suggest a data-driven approach, in which deep learning is used to motivate and enhance the application of more straightforward data processing techniques, as a means for producing results which can be clearly interpreted, easily quantified, and reproducible on generalized datasets.  In practice, the wide availability of technical literature, programming tools, and step-by-step tutorials simultaneously makes machine learning accessible to a wide audience, while obscuring the fact that application to specific datasets requires an understanding of unique, meaningful data features, and of how models can be harnessed to give usable and meaningful analyses.  While common in the field of computer vision, in practice many of the techniques we discuss are added to a machine learning model as a black box, with little understanding of their direct effects on model performance.  Framing deep-learning challenges in the light of real physical systems, we propose means both for thoughtful model design, and for an application of machine learning where the learned features can be visualized and understood by the user.  In this way, analysis of data from high-throughput in situ experiments can become feasible.

\section {Methods}
\subsection{Sample Preparation}
An approximately 1nm Au film was deposited by electron beam assisted deposition in Kurt J. Lesker PVD 75 vacuum deposition system to form nanoparticles with an approximate diameter of 5 nm. The film was directly deposited onto DENSsolutions Wildfire series chips with SiN support suitable for in-situ TEM heating experiments. 

\subsection{TEM Imaging}
Samples were imaged in an FEI Titan 80-300 S/TEM environmental TEM (ETEM) operated at 300kV. Film evolution was studied in vacuum (TEM column base pressure 2x10-7 Torr) at 950$^{\circ}$C. High frame rate image capture utilized a Gatan K2-IS direct electron detector camera at 400 frames per second. Selected images (Figures 2 and S3) were acquired on a JEOL F200 S/TEM operated at 200kV, with images collected on a Gatan OneView camera.

\subsection{Automated Training Set Generation}
\begin{figure}[h]
    \centering
    \includegraphics{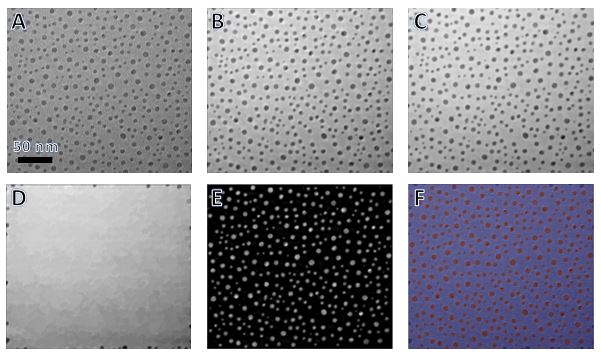}
    \caption{The process of creating a labeled image from a raw image. a.) Example of a raw image. b.) Application of a Gaussian filter for smoothing. Morphological reconstruction by dilation (c), and erosion (d) to extract background features. e.) Image d suctracted from image c. f.) Otsu threshold applied to e (blue/red color scale) overlaid on original image to verify accuracy.}
    \label{automated_processing_fig}
\end{figure}

Raw ETEM images are processed using a series of Gaussian filters, Sobel filters, morphological opening and closing, and thresholding algorithms to produce pseudo-labelled training images (see provided code for reproducing specifics).  All operations are features of the SciKit Image python package\cite{VanderWalt2014}.  As a note, we specify that our dataset is pseudo-labelled, because we take automatically labeled images as ground truth, while traditionally labeled data is produced manually by experts in the field.  Parameters for each of these processing steps, such as the width of the Gaussian filter, are chosen empirically, and the same parameters are applied to all images in the dataset.  Depending on the resolution of the image, and the amount of contrast between the nanoparticles and background in the dataset (which determines the number of required processing steps), automated image processing takes between 10 and 30 seconds per image. Segmentation by this method is faster than manual labeling for particle measurement and localization, which would take tens of minutes per image.  Training set accuracy is evaluated by overlaying labels on raw images and visually inspecting the difference, as there is no way to quantitatively check the ground truth.  Examples of processing steps and training data are shown in Figure \ref{automated_processing_fig}.

 A set of training data was made up of full ETEM images (1792x1920 pixels), collected during a single experiment, downsized via interpolation to a resolution of 512x512-pixels.  Additionally, a second training set with 1024x1024-pixel resolution, made by cropping appropriately sized sections from a full 1792x1920 image, was created to study the impact of increasing pixel resolution on image segmentation performance.  In practice, it is important to consider artifacts introduced by resizing images; stretching or compressing images through interpolation/extrapolation may change local signal patterns.  Cropping sections of images maintains the scale of features in as-collected images, meaning that a model could potentially be trained on many small images (requiring less GPU memory), and then directly evaluated on full images since convolution neural networks do not require specific input/output sizes once training is complete.

\subsection{Programming and Training Machine Learning Models}
All programming was done in Python, with machine learning aspects using the PyTorch framework\cite{Pfeiffer2007}.  The final dataset consisted of 2400 1024x1024-pixel images, which was randomly split into training (70\%, or 1680 images) and test (30\%, or 720 images) sets.  In order to avoid inherent bias due to strong correlation between training and test sets in randomly split consecutive images, a third validation set, collected at a different time but under the same conditions, should be included; we neglect to use this extra dataset, as we only work to show trends in performance as a function of CNN architecture.  All CNNs used rectified linear unit (ReLU) activation after each convolutional layer (except where noted later), the Adam optimizer, and Cross Entropy Loss functions\cite{Kingma2014,Hinton2010}. Since Cross Entropy Loss in PyTorch includes a final softmax activation, a softmax layer was applied to model outputs for inference.  All models were trained for 25 epochs on our System76 Thelio Major workstation using four Nvidia GeForce RTX 2080Ti GPUs, with each model taking 1-2 hours to train.  We note that longer training periods may be required; we used this time frame to make experimentation with network architecture, data pre-processing, and hyper-parameter tuning more feasible in-house. We gauge that models were stable in this training time by tracking loss as a function of epoch number and seeing general convergence, as shown in Supplemental Figure S1. The binary segmentation map which classifies individual pixels as particle or background was obtained by thresholding predicted softmax output for each pixel.

To obtain qualitative data on the particles themselves, both the training set and CNN segmentation output were processed by a connected components algorithm to produce a labeled image which groups pixels into particle regions from which properties such as size and position can be extracted.  This labeling, performed on a binary image, generally takes only one second or less per image
Our base UNet-type architecture for segmenting 512x512 images consisted of three convolutional layers with Max Pooling or Up-sampling (where applicable) on both downscaling and upscaling sides\cite{Ronneberger2015,Scherer2010}. The base model for 1024x1024 images adds an additional level of convolutional layers to each side of the model.  Adding convolutional layers, as described later to increase segmentation accuracy, refers to adding a successive convolutional layer after each down-/up-sampling level of a UNet-type architecture.  Supplemental Figure S2 shows a representation of the CNN architecture used here.

\subsection{Code Availability}
Python code for training image generation, UNet training, and evaluation of results are available at \url{https://github.com/jhorwath/CNN_for_TEM_Segmentation}.

 \section{Data Availability}
 Contact the corresponding author with requestx to view raw data. Sample image sets and all python code used are publicly available on the author's github repository (link provided above).

 \section{Acknowledgements}
 J.P.H and E.A.S acknowledge support through the National Science Foundation, Division of Materials Research, Metals and Metallic Nanostructures Program under Grant 1809398. This research used resources of the Center for Functional Nanomaterials, which is a U.S. DOE Office of Science Facility, at Brookhaven National Laboratory under Contract No. DE-SC0012704, which also provided support to D.N.Z.  The data acquisition was initially supported under Laboratory Directed Research and Development funding at Brookhaven National Laboratory.  The authors thank Yuwei Lin and Shinjae Yoo from Brookhaven National Laboratory for their insights and comments on the manuscript.

 \section{Author Contributions}
 E.A.S, D.N.Z, and R.M. conceived of the ideas for data analysis and experimentation.  D.N.Z collected TEM images with minor contributions from J.P.H, and computational experiments were performed by J.P.H with guidance from R.M.  All authors contributed to preparing the final manuscript.


\bibliographystyle{unsrt}  
\bibliography{references}  
\pagebreak
\section{Supplemental Figures}
\begin{figure}[h]
    \centering
    \includegraphics{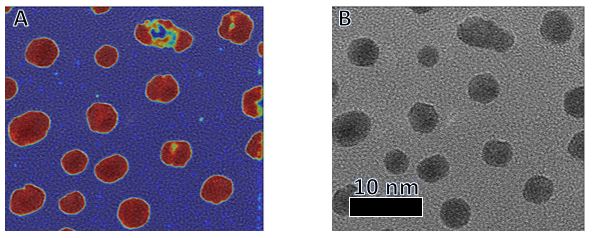}
    \caption{Images showing the errors in identification for large particles in a 512x512-resolution image.  While most particles are correctly labeled, the interior of the largest are missed. a) shows the CNN output overlaid on the raw image, while b) shows the raw image for reference.}
    \label{donut_particles_fig}
\end{figure}

\begin{figure}[h]
    \centering
    \includegraphics{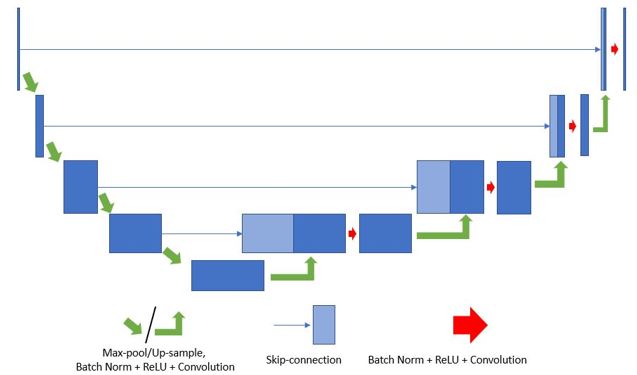}
    \caption{Schematic representation of the UNet-type architecture used on 1024x1024 images.  The red arrow and following blue box are only used in models with a second convolutional layer, as described in the text.}
    \label{UNet_schematic_fig}
\end{figure}

\begin{figure}
    \centering
    \includegraphics{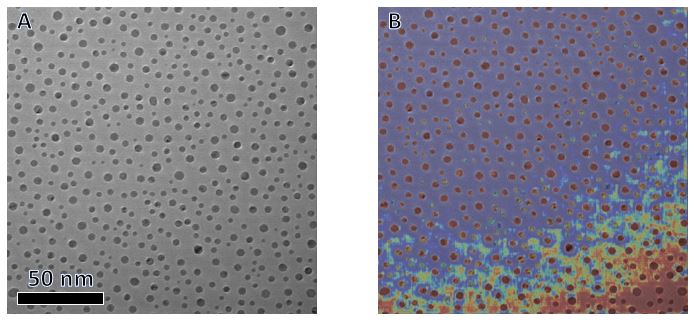}
    \caption{a) Raw image which represents the average of 40 consecutive frames.  Though averaging images smooths the background making particle boundaries more clear to the human eye, the poor segmentation in the bottom right corner of b) shows that a neural network trained on noisy data is not effective on cleaner data.}
    \label{averaged_images_fig}
\end{figure}

\begin{figure}[h]
    \centering
    \includegraphics{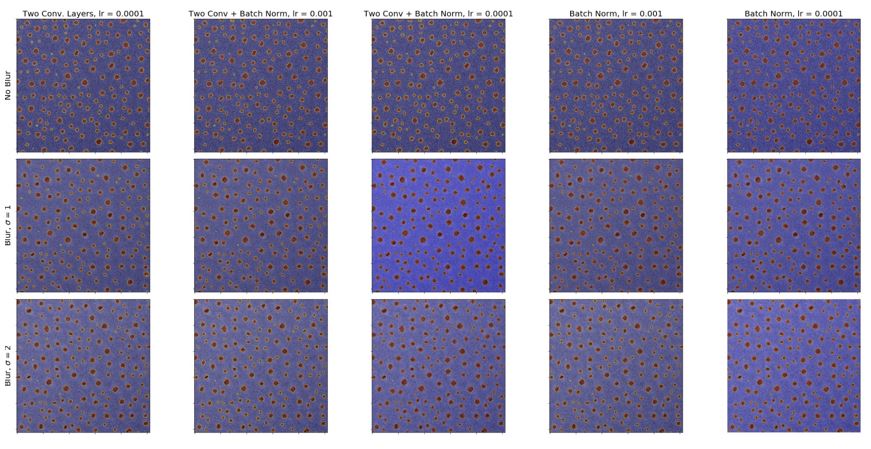}
    \caption{Results of the 15 models tested to improve performance on high resolution images.}
    \label{15_models_fig}
\end{figure}

\begin{figure}[h]
    \centering
    \includegraphics{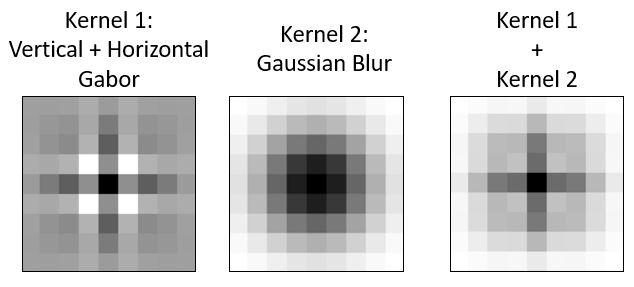}
    \caption{Schematic example showing that the sum of a horizontal Gabor filter, a vertical Gabor filter, and Gaussian blur produces a kernel similar to that learned by our simple one-layer CNN.}
    \label{simulate_gabor_fig}
\end{figure}

\begin{figure}[h]
    \centering
    \includegraphics{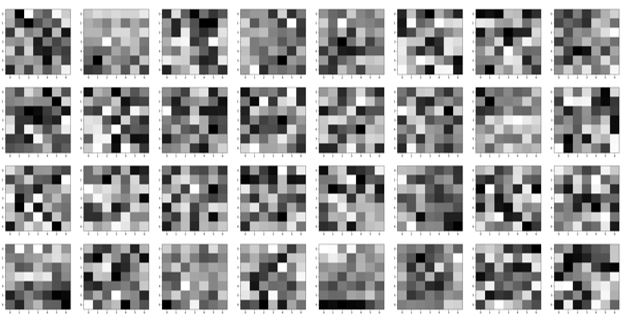}
    \caption{Visualization of all learned filters in the CNN consisting of one layer with 32 filters.}
    \label{32_filters_fig}
\end{figure}

\begin{figure}[h]
    \centering
    \includegraphics{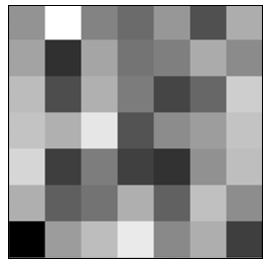}
    \caption{Mean of all 32 convolutional kernels shown in Figure \ref{32_filters_fig}.}
    \label{mean_kenel_fig}
\end{figure}

\begin{figure}[h]
    \centering
    \includegraphics{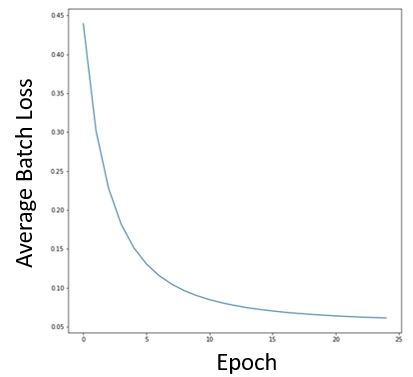}
    \caption{Plot showing convergence of model loss as a function of training time (number of epochs)}

    \label{loss_convergence_fig}
\end{figure}


\end{document}